# Nanotube-Based NEMS: Control vs. Thermodynamic Fluctuations

Olga V. Ershova[*] and Irina V. Lebedeva[†]

*Moscow Institute of Physics and Technology, 141701, Dolgoprudny, Moscow Region, Russia*

Yurii E. Lozovik[‡] and Andrey M. Popov[§]

*Institute of Spectroscopy, 142190, Troitsk, Moscow region, Russia*

Andrey A. Knizhnik[**] and Boris V. Potapkin

*RRC "Kurchatov Institute", 123182, Moscow, Russia*

*Kintech Lab Ltd, 123182, Moscow, Russia*

Oleg N. Bubel, Eugene F. Kislyakov and Nikolai A. Poklonski

*Belarusian State University, 220030, Minsk, Belarus*

**ABSTRACT**

Multi-scale simulations of nanotube-based nanoelectromechanical systems (NEMS) controlled by a nonuniform electric field are performed by an example of a gigahertz oscillator. Using molecular dynamics simulations, we obtain the friction coefficients and characteristics of the thermal noise associated with the relative motion of the nanotube walls. These results are used in a phenomenological one-dimensional oscillator model. The analysis based both on this model and the Fokker–Planck equation for the oscillation energy distribution function shows how thermodynamic fluctuations restrict the possibility of controlling NEMS operation for systems of small sizes. The parameters of the force for which control of the oscillator operation is possible are determined.

---

[*] olga.v.ershova@gmail.com

[†] lebedeva@kintech.ru

[‡] lozovik@isan.troitsk.ru

[§] am-popov@isan.troitsk.ru

[**] knizhnik@kintech.ru

**PACS** numbers: 61.46Fg, 85.85+j

## I. INTRODUCTION

The ability of the free relative sliding and rotation of carbon nanotube walls[1,2,3] and their excellent "wearproof" characteristics allow the use of carbon nanotube walls as movable elements in nanoelectromechanical systems (NEMS). A number of devices offering great promise for applications in NEMS and based on the use of the relative motion of carbon nanotube walls have been proposed recently. These devices include rotational[4,5] and plain nanobearings, nanogears[6], electromechanical nanoswitchs[7], nanoactuators[8,9], Brownian motors[10], nanobolt-nanonut pairs[11,12,13], and gigahertz oscillators[14,15]. Furthermore, nanomotors based on the relative rotation of carbon nanotube walls[16,17,18,19] and memory cells based on the relative sliding of carbon nanotube walls[20] have been implemented.

The crucial issue in nanotechnology is the actuation of the NEMS components in a controllable way. A new method for controlling the motion of NEMS based on carbon nanotubes[21] was proposed lately. Namely, the wall of a nanotube has an electric dipole moment if electron donors or/and acceptors are adsorbed at the ends of the wall. The motion of such a functionalized wall with an electric dipole moment can be controlled by a nonuniform electric field. Here, molecular dynamics (MD) simulations are performed to demonstrate the feasibility of this control method.

As compared to microelectromechanical systems, the principal feature of NEMS related to a small number of atoms in the system is the significance of thermodynamic fluctuations in NEMS. Here, we perform multi-scale simulations to study the influence of these fluctuations on the NEMS operation by the example of a nanotube-based gigahertz oscillator. We believe that the approach developed and the results of this study can be also useful for other types of NEMS, such as artificial molecular rotors[22] and so on.

The gigahertz oscillator based on a double-walled carbon nanotube is one of the simplest nanotube-based NEMS. Thus, it is commonly used as a model system for studying the tribological behavior of the nanotube-based NEMS[23,24,25,26,27,28,29,30,31,32,33,34,35,36,37] and possible methods for controlling their operation[38,39]. The scheme, operational principles, and theory of the gigahertz oscillator based on the relative sliding of carbon nanotube walls were considered by Zheng *et al.*[14,15]. The oscillator was proposed to be used as a part of the device for surface profiling[40]. Upon the telescopic extension of the inner wall outside the outer wall, the van der Waals force $F_W$ turns the inner wall back into the outer wall, thereby makes this NEMS oscillate. However, such oscillations are anharmonic and dissipative with the Q-factor $Q \approx 10-1000$[24,26,27,28,32] (Q-factor of the system is the ratio of the total oscillation energy to the energy loss per one oscillation period). The frequency of damping oscillations increases with decreasing oscillation amplitude. Thus, this frequency increases with time[27,30]. Consequently, to provide the stationary operation of the gigahertz oscillator, that is, to keep its frequency constant, it is necessary to compensate the energy dissipation by the work of an external force. To obtain the critical value of this force (i.e. the minimum value at which the work of the external force can compensate the energy dissipation), the energy balance in the controlled operation of the gigahertz oscillator was analyzed. It was shown that the critical amplitude $F_{0c}$ of the harmonic control force $F(t) = F_0 \cos(2\pi t / T_s)$ applied to the movable inner wall attains a minimum value for the oscillator with walls of equal lengths and is given by the expression

$$F_{0c} = \frac{\pi^2 F_W}{32Q}, \tag{1}$$

where $Q$ is the Q-factor corresponding to the oscillation with period $T_s$.

Here we study the characteristics of the actuation of a functionalized nanotube wall by a nonuniform electric field through multi-scale simulations of the controlled operation of the (5,5)@(10,10) nanotube-based gigahertz oscillator. Molecular dynamics (MD) simulations are performed to demonstrate the feasibility of this control method. To estimate the critical amplitude of the control force, we calculate the Q-factor of the gigahertz oscillator using MD simulations of the damping oscillations. However, the MD technique does not allow studying systems containing a large number of atoms within a long simulation time. So, the operation of the gigahertz oscillator is analyzed within the framework of a phenomenological one-dimensional model with the parameters derived from MD simulations. This analysis is used to investigate the system behavior in the course of its relaxation to the stationary operation and to determine the conditions under which the stationary operation of this NEMS is possible.

By now, thermodynamic fluctuations in NEMS were considered and investigated using MD simulations only to demonstrate the possibility of the relative diffusion of the NEMS components[41] and regarding the equilibrium rotational[42] and translational dynamics of walls in carbon nanotubes. As for the nanotube-based NEMS, such diffusion can be used in Brownian motors. However, diffusion[43] or displacement of the NEMS components due to thermodynamic fluctuations can disturb the NEMS operation[9,43]. For example, it was shown that thermodynamic fluctuations restrict the minimum sizes of the electromechanical nanothermometer based on the interaction of nanotube walls and the electromechanical nanorelay based on the relative motion of nanotube walls. The MD simulations performed here revealed substantial fluctuations in the nanotube-based gigahertz oscillator. The analysis of the motion equation of this NEMS indicates the critical influence of these fluctuations on the possibility of controlling its operation. In this way, the principal restrictions imposed by thermodynamic fluctuations on the possibility of controlling the NEMS operation are studied within the framework of the phenomenological one-dimensional model and analyzed on the basis of the Fokker–Planck equation for the oscillation energy distribution function.

The paper is organized in the following way. The results of the MD simulations of the operation of the gigahertz oscillator are given in Sec. II. In Sec. III, we examine the influence of the characteristics of the control force on the operation of the gigahertz oscillator within the framework of the phenomenological model with the parameters derived from the MD simulations. The effect of thermodynamic fluctuations on the NEMS operation is analyzed in Sec. IV. Our conclusions are summarized in Sec. V.

## II. MOLECULAR DYNAMICS SIMULATION OF GIGAHERTZ OSCILLATOR

To obtain the oscillator Q-factor, study the characteristics of the thermal noise, and demonstrate the possibility of controlling the motion of a functionalized nanotube wall by a nonuniform electric field, we performed MD simulations of the free and controlled operation of the (5,5)@(10,10) nanotube-based oscillator (see FIG. 1). For this double-walled nanotube, there is no resonance between the telescopic oscillation and other nanotube vibrations, and the populations of the vibrational levels are in thermal equilibrium. Both nanotube walls were taken equal to 3.1 nm in length. One end of the inner wall was capped and the other end was open and terminated with hydrogen atoms (see FIG. 1). Both ends of the outer wall were open and not functionalized. The charge distribution in the inner wall was calculated by the semiempirical method of molecular orbitals with the PM3 parameterization of the Hamiltonian[44]. The adequacy of the PM3 parameterization in the case under consideration was demonstrated by calculations of bond lengths in the $C_{60}$ fullerene with the symmetry $I_h$: the calculated bond lengths agree with their experimental counterparts within $10^{-4}$ nm[45].

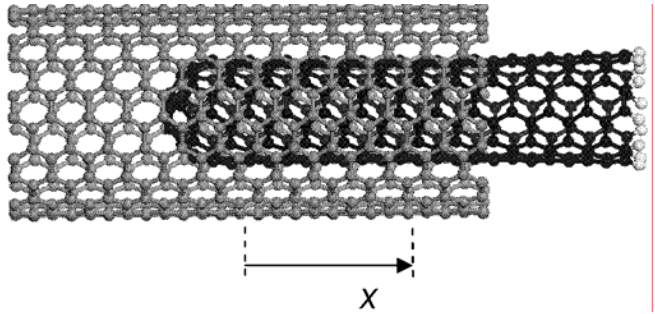

FIG. 1. Gigahertz oscillator based on the (5,5)@(10,10) double-walled nanotube. Hydrogen atoms are shown in light gray.

The analysis of the free and controlled behavior of the gigahertz oscillator was performed using empirical interatomic potentials. Interaction between the inner and outer wall atoms $i$ and $j$ at distance $r_{ij}$ was described by the Lennard–Jones 12–6 potential

$$V_{\text{LJ}}(r_{ij}) = 4\varepsilon\left(\left(\frac{\sigma}{r_{ij}}\right)^{12} - \left(\frac{\sigma}{r_{ij}}\right)^{6}\right) \qquad (2)$$

with the parameters $\varepsilon_{\text{CC}} = 3.73$ meV, $\sigma_{\text{CC}} = 3.40$ Å and $\varepsilon_{\text{CH}} = 0.65$ meV, $\sigma_{\text{CH}} = 2.59$ Å for the carbon-carbon and carbon-hydrogen interaction, respectively, taken from the AMBER database[46] for aromatic carbon and hydrogen bonded to aromatic carbon. The parameters provide a consistent description of the pairwise carbon-carbon and carbon-hydrogen interactions. The cut-off distance of the Lennard–Jones potential was taken equal to 12 Å. The covalent carbon-carbon and carbon-hydrogen interactions inside the walls were described by the empirical Brenner potential[47], which was shown to correctly reproduce the vibrational spectra of carbon nanotubes[48]. So the Hamiltonian of the oscillator based on a double-walled nanotube was given by

$$H = K^{(\text{out})} + K^{(\text{in})} + V_{\text{B}}^{(\text{out})} + V_{\text{B}}^{(\text{in})} + \sum_{i,j} V_{\text{LJ}}(r_{ij}), \qquad (3)$$

where $K^{(\text{out})}$ and $K^{(\text{in})}$ are the total kinetic energies of the inner and outer walls, respectively, $V_{\text{B}}^{(\text{out})}$ and $V_{\text{B}}^{(\text{in})}$ are the total Brenner potentials of the walls, and the last term is the sum over all pairs of atoms $i$ of the inner wall and atoms $j$ of the outer wall.

The value of the total van der Waals force, which retracts the telescopically extended inner wall back into the outer wall, was found to be $F_{\text{W}} = 1180$ pN for the above parameters of the Lennard–Jones potential. For large-diameter nanotubes, the interwall van der Waals energy should be proportional to the overlap area between the walls. Therefore, the total van der Waals force is proportional to the nanotube diameter. The van der Waals force per unit nanotube diameter was found to be about 0.1 – 0.2 N/m in the experiments of Cumings et. al. and Kis et al.. We obtained the value of the van der Waals force per unit nanotube diameter of about 0.3 N/m. However, one should take into account that the measurements[2,3] were performed for large-diameter multi-walled nanotubes, whereas a non-linear dependence of the van der Waals force on the nanotube diameter should be expected for small-diameter nanotubes. The barrier to relative sliding of the walls was calculated to be 0.008 meV per atom of the outer wall, in agreement with the recent result obtained for the (5,5)@(10,10) nanotube through LDA calculations[49]. The Lennard-Jones potential was also shown to provide the interlayer interaction energy in graphite of about 62 meV/atom, which is consistent with the experimental value 52±5 meV/atom obtained recently from the experiments on the thermal desorption of polyaromatic hydrocarbons from a graphitic surface[50] but greater than the values reported earlier (see Ref. 51 and references therein).

An in-house MD-kMC code[52] was implemented. The code used the velocity Verlet algorithm and neighbor lists to improve the computing performance. The time step was 0.2 fs, which is about two orders of magnitude shorter than the period of thermal vibrations of hydrogen atoms. The initial configuration of the nanotube was optimized at zero temperature. The initial velocities of the atoms were distributed according to the Maxwell-Boltzmann distribution with a doubled pre-heating temperature $2T_0$. This provided that during the MD simulation, the initial distribution of velocities reached the Maxwell-Boltzmann temperature corresponding to the pre-heating temperature $T_0$ in less than 0.02 ps. To start the oscillation, the inner wall was pulled out along the nanotube axis at the distance $s = 1$ nm (about 30% of its length) and released with the zero center-of-mass velocity. The outer wall was fixed at three atoms. The relative fluctuations of the total energy of the system caused by numerical errors were less than 0.3% of the interwall van der Waals energy.

To get the temperature dependence of the oscillator Q-factor, we performed microcanonical MD simulations of free oscillations (see FIG. 2) at pre-heating temperatures of 0, 50, 100, 150 and 300 K. The oscillation energy $E$ is given by the sum of the center-of-mass kinetic energy of the movable wall and the excess of the interwall van der Waals energy $V(x)$ associated with the displacement $x$ of the movable wall from the position corresponding to the minimum of the interwall van der Waals energy ($V(0) = 0$)

$$E = V(x) + \frac{m\upsilon^2}{2}, \qquad (4)$$

where $m$ is the mass of the movable wall, and $\upsilon$ is its center-of-mass velocity. The calculation of the oscillation energy was performed at the moments when the movable wall crossed the position $x = 0$ corresponding to the minimum of the interwall van der Waals energy. So the oscillation energy was found at these moments as $E = mv^2/2$. Since the oscillation energy loss $\Delta E$ over a half period of the oscillation is much smaller than the oscillation energy $E$ ($\Delta E << E$), the Q-factor was estimated using the ratio $\Delta E / E$ averaged over a single simulation run

$$Q = \left\langle \frac{2\Delta E}{E} \right\rangle^{-1}. \tag{5}$$

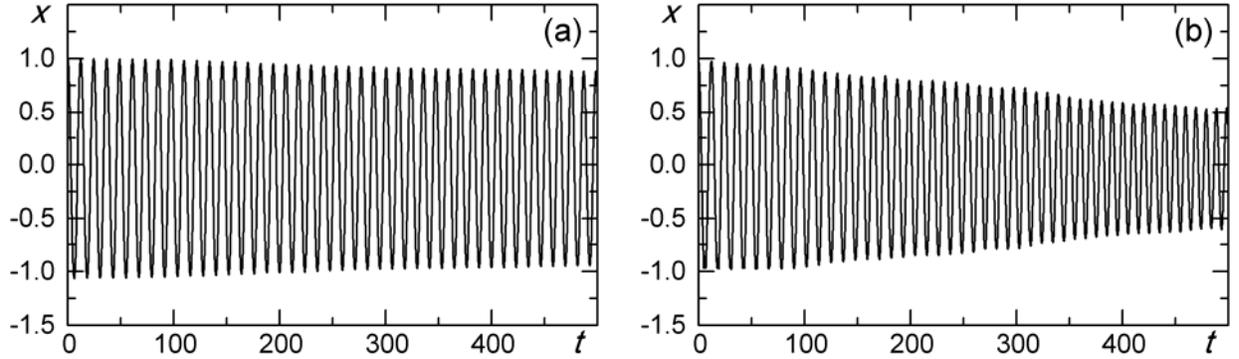

FIG. 2. Calculated displacement of the free oscillating movable wall $x$ (in nm) as a function of time $t$ (in ps) at pre-heating temperatures of 50 (a) and 300 K (b).

There are two contributions to the Q-factor calculation error. On the one hand, the Q-factor depends on the oscillation amplitude. Thus, the damping of the oscillation leads to a change in the Q-factor. This effect becomes relevant at a long simulation time. On the other hand, there is also a stochastic contribution to the Q-factor calculation error related to thermodynamic fluctuations, which is high at a short simulation time and decreases with its increase. We took a simulation time of 500 ps, which minimizes the Q-factor calculation error in a single run and provides the accuracy of the Q-factor calculations within 20%.

Temperature $T$ of the system was monitored based on the total thermal kinetic energy of the atoms in the system

$$\frac{3}{2}k_B TN = \left\langle K^{(out)} + K^{(in)} - \frac{mv^2}{2} \right\rangle_{\Delta t}, \tag{6}$$

where $k_\text{B}$ is Boltzmann's constant, $N$ is the number of atoms in the system. The total thermal kinetic energy was averaged over a time interval $\Delta t = 1$ ps, as this time interval must be much longer than the period of thermal vibrations (about 0.1 ps) and, on the other hand, much shorter than the period of the telescopic oscillation (about 10 ps). The temperature change over the simulation time was less than 9% at pre-heating temperatures of 50, 100, 150 and 300 K. At a pre-heating temperature of 0 K, temperature increased to 2 K within the simulation time. An estimation[24] showed that the influence of quantum effects on the dissipation rate can be neglected at temperatures above 10 K. The calculated Q-factor $Q$ at different pre-heating temperatures is given in TABLE I.

TABLE I. Calculated Q-factor $Q$, relative deviation $\delta_E$ of the relative energy change $\Delta E / E$ over the oscillation half-period, and critical amplitudes of the control force $F_{0c}$ and voltage* $U_{0c}$ at different preheating temperatures $T_0$.

| $T_0$ (K) | $Q$ | $\delta_E$ | Long nanotube case | | Short nanotube case | |
|---|---|---|---|---|---|---|
| | | | $F_{0c}$ (pN) | $U_{0c}$ (V) | $F_{0c}$ (pN) | $U_{0c}$ (V) |
| 0 | 700 ± 350 | 4.4 | 0.52 | 6.0 | 0.40 | 4.7 |
| 50 | 250 ± 50 | 1.2 | 1.4 | 17 | 1.1 | 13 |
| 100 | 160 ± 30 | 1.4 | 2.3 | 26 | 1.8 | 20 |
| 150 | 140 ± 30 | 1.5 | 2.7 | 31 | 2.1 | 24 |
| 300 | 55 ± 11 | 1.5 | 6.6 | 77 | 5.2 | 60 |

*Voltage is calculated for a spherical capacitor with plate radii of 100 and 110 nm.

To estimate the level of fluctuations in the system, we also calculated the root-mean-square deviation of the relative energy change $\Delta E / E$ over every half period of the oscillation. Significant fluctuations of the quantity $\Delta E / E$ were revealed (see FIG. 3). The calculated relative root-mean-square deviation $\delta_E$ of $\Delta E / E$ at different temperatures is given in TABLE I.

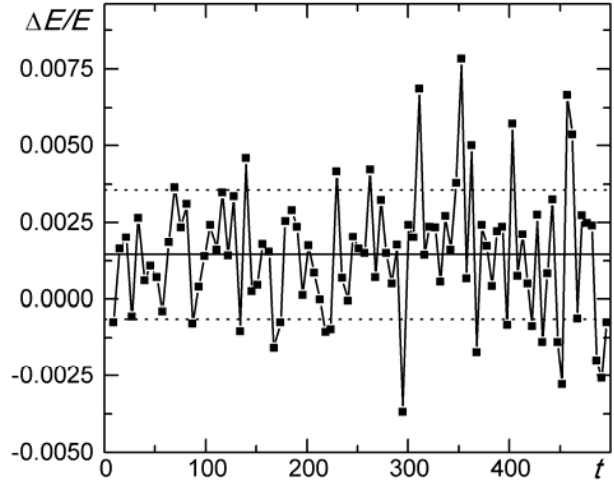

FIG. 3. Calculated relative loss $\Delta E / E$ of the oscillation energy over a half-period of the oscillation as a function of time $t$ (in ps) at a temperature of 300 K. The average value and the root-mean-square deviation are shown with the solid and dotted lines, respectively.

It was found that the Q-factor of the oscillator $Q$ strongly increases with decreasing temperature (see TABLE I), in agreement with the other papers[24,34,35,36,37]. Nevertheless, the Q-factor does not go to infinity at the zero pre-heating temperature. The relative deviation $\delta_E$ of $\Delta E / E$ weakly depends on temperature in the temperature range of 50 – 300 K. However, the considerable increase in $\delta_E$ is observed for the zero pre-heating temperature. Explanations of these results are presented below.

The investigation of the tribological properties of the (5,5)@(10,10) nanotube-based oscillator has revealed that the populations of the nanotube vibrational levels are in thermal equilibrium, which justifies the applicability of the fluctuation-dissipation theorem[53,54] for this system. As the dissipation rate is proportional to the number of excited phonons, it was shown that the inverse Q-factor should linearly depend on temperature, in agreement with TABLE I. Since a certain energy exchange between the telescopic oscillation of the movable wall and the other degrees of freedom exists at the zero temperature even in the classical limit, the Q-factor remains finite at the zero pre-heating temperature.

According to the analysis performed within the framework of the fluctuation-dissipation theorem[53,54], the relative deviation $\delta_E$ of the relative energy change $\Delta E/E$ over a half-period of the oscillation is determined by the expression

$$\delta_E \approx 2.5\sqrt{\frac{k_B T Q}{E}}. \tag{7}$$

As the Q-factor of the considered gigahertz oscillator is almost inversely proportional to temperature (see TABLE I and also Ref. 24), it follows from Eq. (7) that $\delta_E$ weakly depends on temperature. However, this result is valid unless the applicability conditions of the fluctuation-dissipation theorem are not violated. One of these applicability conditions is that all degrees of freedom except for one should be in thermal equilibrium. We suppose that, at very low temperatures, the system is highly non-equilibrium. This results in the deviation from Eq. (7) and the great value of $\delta_E$ for the zero pre-heating temperature (see TABLE I). Note that, according to Eq. (7), the significant fluctuations of the relative energy change $\Delta E/E$ over a half period of the oscillation should be attributed to the small oscillation energy, i.e. the small size of the system under consideration or the small amplitude.

In the simulations of the controlled oscillations, the hydrogen functionalized inner wall was exposed to the harmonic electric field of a spherical capacitor. In this case, one more term was added to the Hamiltonian (3) of the system

$$V_{ex} = \sum_i q_i \phi(\vec{r}_i), \tag{8}$$

where $q_i$ is the charge of the atom $i$ of the movable wall, and $\phi(\vec{r}_i)$ is electric field potential at the position of the atom $i$ with the radius-vector $\vec{r}_i$ from the center of the spherical capacitor

$$\phi(\vec{r}_i) = -\frac{R_1 R_2}{R_2 - R_1} \frac{1}{|\vec{r}_i|} U_0 \cos(\omega t + \varphi). \tag{9}$$

Here $U_0$ is the amplitude of the applied voltage, $\omega$ is the angular frequency of the electric field, $\varphi$ is the initial phase shift between the electric field and the velocity of the movable wall, $R_1$ and $R_2$ are the radii of the inner and outer plates of the capacitor, respectively. The angular frequency of the electric field equaled the oscillation angular frequency $\omega_0$ corresponding to the initial oscillation amplitude $s = 1$ nm ($\omega = \omega_0$). The initial phase shift between the electric field and the inner wall velocity equaled zero ($\varphi = 0$). The radii of the plates of the spherical capacitor were taken equal to $R_1 = 100$ nm and $R_2 = 110$ nm.

Furthermore, in these simulations, the temperature of the outer wall was maintained by periodically rescaling atomic velocities every 0.1 ps (the Berendsen thermostat[55]). Though only the outer wall was kept in contact with the thermostat, our estimation demonstrates that the temperatures of both walls are almost equal. In fact, the dissipation rate at a temperature of 300 K was found to be $W \approx 10^{-9}$ W. The thermal conductivity between the nanotube walls can be assumed equal to that between graphite layers[56] $\chi_\perp \approx 5$ W/(m·K). Under stationary conditions, the dissipation rate should be equal the rate of heat transfer between the walls. Supposing that the heat flux is identical through all coaxial cylindrical surfaces between the inner and outer walls, one gets the temperature difference between the nanotube walls

$$\Delta T = \frac{W}{2\pi L \chi_\perp} \ln\left(1 + \frac{\delta R}{R_{in}}\right) \approx 10^{-2} \text{K}. \qquad (10)$$

Here, $L = 3.1$ nm is the wall length, $\delta R = 0.34$ nm is the interwall distance, $R_{in} = 0.34$ nm is the inner wall radius. As it is seen, the temperature difference between the walls is negligible.

The results of the MD simulations of the controlled oscillations are presented in FIG. 4. The amplitude of the applied voltage was 60.5 and 61.4 V at a temperature of 300 K and 10.9 V at 50 K. The oscillations shown in FIG. 4 correspond to the control force amplitude greater than the critical value for the considered oscillator. In this case, the variation of the oscillation amplitude is observed, in agreement with the results obtained below using the phenomenological one-dimensional model.

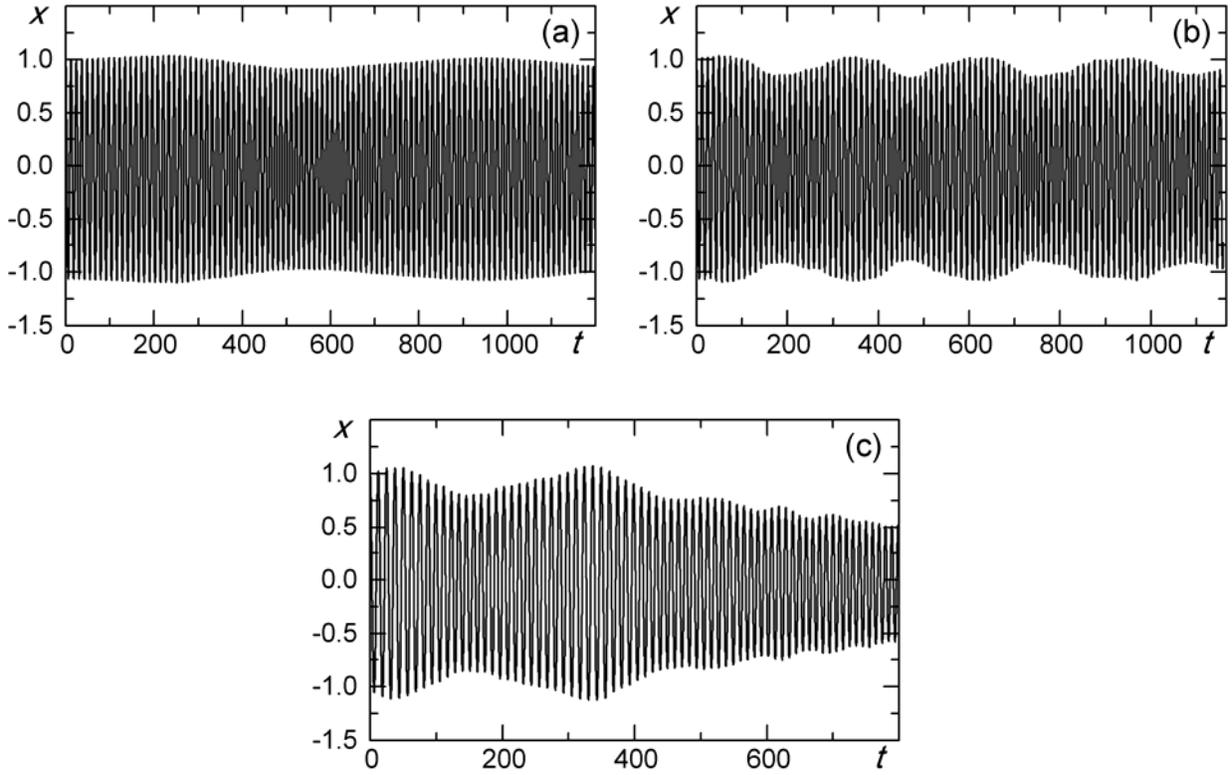

FIG. 4. Calculated displacement of the movable wall $x$ (in nm) for the controlled oscillations as a function of time $t$ (in ps) at temperatures of 50 (a) and 300 K (b, c). The voltage amplitude is (a) 10.9 V; (b) 60.5 V; (c) 61.4 V.

Recently, a nanomotor based on the relative rotation of nanotube walls in a multi-walled carbon nanotube was implemented. In this experiment, the voltage applied between the nanotube wall and the control electrode reached 100 V. This value is of the order of the voltage amplitude needed to sustain the oscillation at a constant amplitude in the performed MD simulations. Thus, the MD simulations have demonstrated that the method under consideration can be used to control the motion of the nanotube-based NEMS.

The amplitude of the applied voltage needed to sustain the oscillation at a constant amplitude can be decreased if the dipole moment of the inner wall is increased. This can be achieved, for example, through the adsorption of hydrogen and fluorine atoms at opposite open ends of the nanotube. Moreover, as it was shown above, the Q-factor is inversely proportional to temperature. Therefore, oscillator operation at low temperatures requires a smaller amplitude of the control force. For instance, at the temperature of liquid helium (4.2 K), the necessary amplitude of the applied voltage is only several volts.

It should also be mentioned that, in some cases, even at relatively high amplitudes of the control force, the breakdown of the oscillation occurs (see FIG. 4c). The analysis below shows that this breakdown is induced by thermodynamic fluctuations. In fact, let us suppose that occasionally a significant negative fluctuation of the oscillation energy occurs. This means that the oscillation amplitude decreases. Since the oscillator frequency strongly depends on the oscillation amplitude[14,15], the oscillator gets out of the resonance with the control force. This leads to a decrease of the work of the control force, which, in turn, results in a further decrease in the oscillation energy. As follows from this explanation, the stability of the oscillator operation might be improved with increasing the amplitude of the control force (above the critical value) or with decreasing the level of fluctuations in the system. It is seen that the breakdown of the oscillation induced by thermodynamic fluctuations should be expected for any strongly anharmonic oscillator.

### III. PHENOMENOLOGICAL MODEL OF GIGAHERTZ OSCILLATOR

MD simulations allow the study of system behavior only at times of a few nanoseconds. To reach longer simulation times, a phenomenological model with the parameters derived from MD simulations is proposed. If there is no resonance between the telescopic oscillation and other vibrations in the nanotube (which is the case for the considered (5,5)@(10,10) nanotube), the oscillator dynamics may be roughly described by a one-dimensional equation of motion

$$m\ddot{x}(t) = F_{\text{vdW}}(x(t)) + F_{\text{fr}}(\dot{x}(t)) + \xi(t) + F(t). \qquad (11)$$

Here, $x$ is the displacement of the movable wall; $m$ is its mass; $F_{\text{vdW}}(x)$ is the interwall van der Waals force; $F_{\text{fr}}(\dot{x})$ is the friction force modeling energy dissipation; $\xi(t)$ is the white noise representing random fluctuating forces; and $F(t)$ is the control force. The initial conditions were $x(0) = s = 1\,\text{nm}$, $\dot{x}(0) = 0$.

As above, we considered the harmonic control force $F(t) = F_0 \cos(\omega t + \varphi)$, where $F_0$ is control force amplitude, $\omega$ is the angular frequency of the control force, and $\varphi$ is the initial phase shift between the control force and the velocity of the movable wall. At a relatively low oscillation energy, the dynamic friction force is proportional to the oscillator velocity[36,37] with the friction coefficient $\eta$, $F_{\text{fr}}(\dot{x}(t)) = -\eta \dot{x}(t)$. The friction coefficient $\eta$ was chosen so as to reproduce the Q-factor of the oscillator at the initial oscillation amplitude $s = 1$ nm (see TABLE I). Since the phenomenological model was used here to study the qualitative behavior of the system, the dependence of the friction coefficient $\eta$ on the oscillation amplitude was disregarded, i.e. the friction coefficient $\eta$ was supposed to be constant. To determine the conditions of the stationary operation and study the relaxation of the oscillator to the stationary operation, the thermal noise was neglected ($\xi(t) \equiv 0$).

Two approximations for the interwall van der Waals force $F_{\text{vdW}}(x)$ were considered. The first one was the calculated dependence of the interwall force on the displacement between the walls for the (5,5)@(10,10) nanotube 3.1 nm in length, which was used in our MD simulations (see FIG. 5). To calculate this dependence, the walls were separately relaxed and the inner wall was then rigidly shifted along the wall common axis. The interwall force was found as the partial derivative of the total interwall van der Waals energy with respect to the inner wall displacement. Since the nanotube under consideration is rather short, the region in which the interwall van der Waals force increases from zero to the nearly constant value $F_W \approx 1180$ pN ($x < 0.5$ nm) is comparable to the oscillation amplitude (see FIG. 5). Thus, Eq. (11) with this approximation of the interwall van der Waals force represents a "short nanotube" case of the proposed model. Eq. (11) was solved numerically with a time step of 1 fs, and the friction coefficient $\eta$ for the given temperature was fitted so that the model reproduced the values of the Q-factor obtained in the MD simulations.

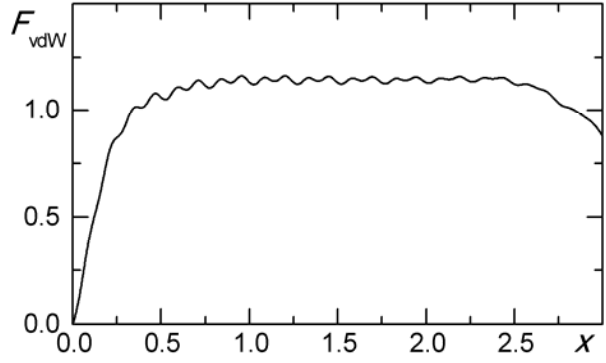

FIG. 5. Calculated interwall van der Waals force $F_{vdW}$ (in nN) of the (5,5)@(10,10) nanotube 3.1 nm in length as a function of the displacement of the movable wall $x$ (in nm).

For a long nanotube, one can neglect the short region $x < 0.5$ nm in the dependence of the interwall force on the displacement between the walls, where the force increases from zero to the nearly constant value. Thus, the interwall van der Waals force $F_{vdW}(x)$ can be assumed constant[14,15]

$$F_{vdW}(x) = -F_W \text{sign}(x). \qquad (12)$$

The magnitude of the interwall force $F_W$ was taken equal to that in the "short nanotube" case $F_W = 1180$ pN (see FIG. 5). Approximation (12) of the interwall van der Waals force provides the following expression for the friction coefficient

$$\eta = \frac{3}{8Q}\sqrt{\frac{F_W m}{2s}} = \frac{3m}{2QT_s}, \qquad (13)$$

where

$$T_s = 4\sqrt{\frac{2ms}{F_W}} \qquad (14)$$

is oscillation period corresponding to amplitude $s$.

In this "long nanotube" case, motion equation (11) can be solved semi-analytically with the time step equal to the oscillation half-period. This leads to the decrease in the calculation time by a factor of 10 – 100 compared to that in the "short nanotube" case.

In the "long nanotube" case, the critical amplitude of the control force is determined by Eq. (1). In the "short nanotube" case, the interwall van der Waals force is not constant and has a smaller average value. So, in the latter case the critical amplitude of the control force is less than the value given by Eq. (1). The calculated critical amplitudes of the control force and voltage for the "short nanotube" case are presented in TABLE I. The critical voltage amplitudes agree with the results of the MD simulations (see FIG. 4a and FIG. 4b) for the same nanotube within the accuracy of the Q-factor calculations.

Let us consider the conditions of switching the control force for the free oscillation with an amplitude $s = 1\,\text{nm}$. The possibility of the stationary operation mode for the given $F_0 / F_{0c}$ ratio is determined by the initial phase shift $\varphi$ between the control force and relative velocity of the walls and the angular frequency $\omega$ of the control force. We examined the ranges of normalized parameters of the control force $F_0 / F_{0c}$, $\omega / \omega_0$ (where $\omega_0$ is the angular frequency corresponding to the oscillation with the amplitude $s = 1\,\text{nm}$) and $\varphi / \pi$ where the stationary mode is possible for both cases of the proposed model. As it can be seen in FIG. 6 and FIG. 7, on increasing the amplitude of the control force $F_0$ or decreasing the Q-factor, the ranges of $\omega$ and $\varphi$ which correspond to the stationary operation mode become wider. At the high amplitude of the control force $F_0 = 10 - 100 F_{0c}$, switching of the control force is reasonable almost at an arbitrary moment (see FIG. 6 and FIG. 7).

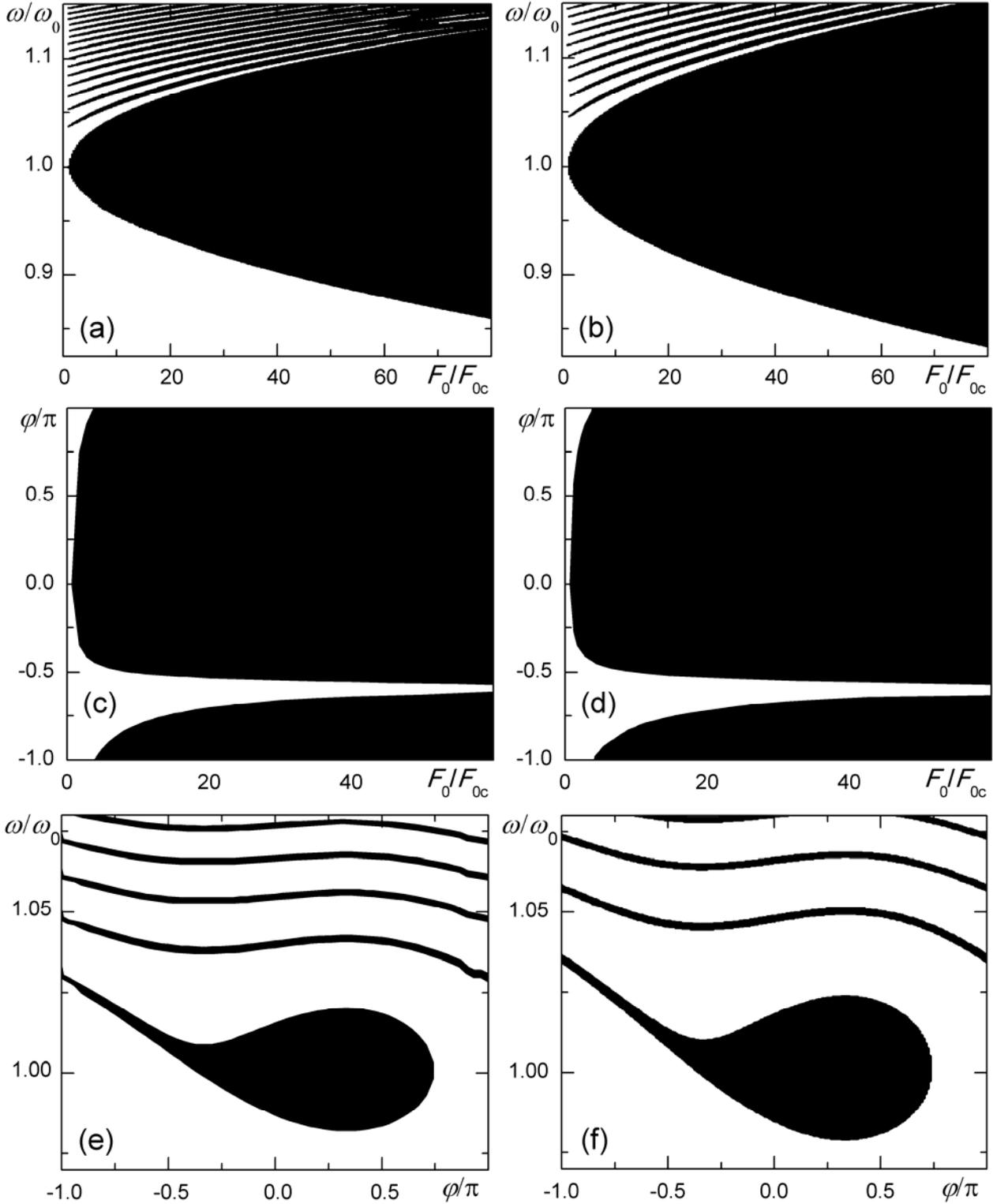

FIG. 6. Calculated parameters of the control force $F_0/F_{0c}$, $\omega/\omega_0$ and $\varphi$ (shown in black) for which the stationary operation mode is possible at $Q = 500$. (a), (c), (e) "Short nanotube" case, (b), (d), (f) "long nanotube" case of the phenomenological model. (a), (b) $\varphi = 0$; (c), (d) $\omega/\omega_0 = 1$; (e), (f) $F_0/F_{0c} = 2$.

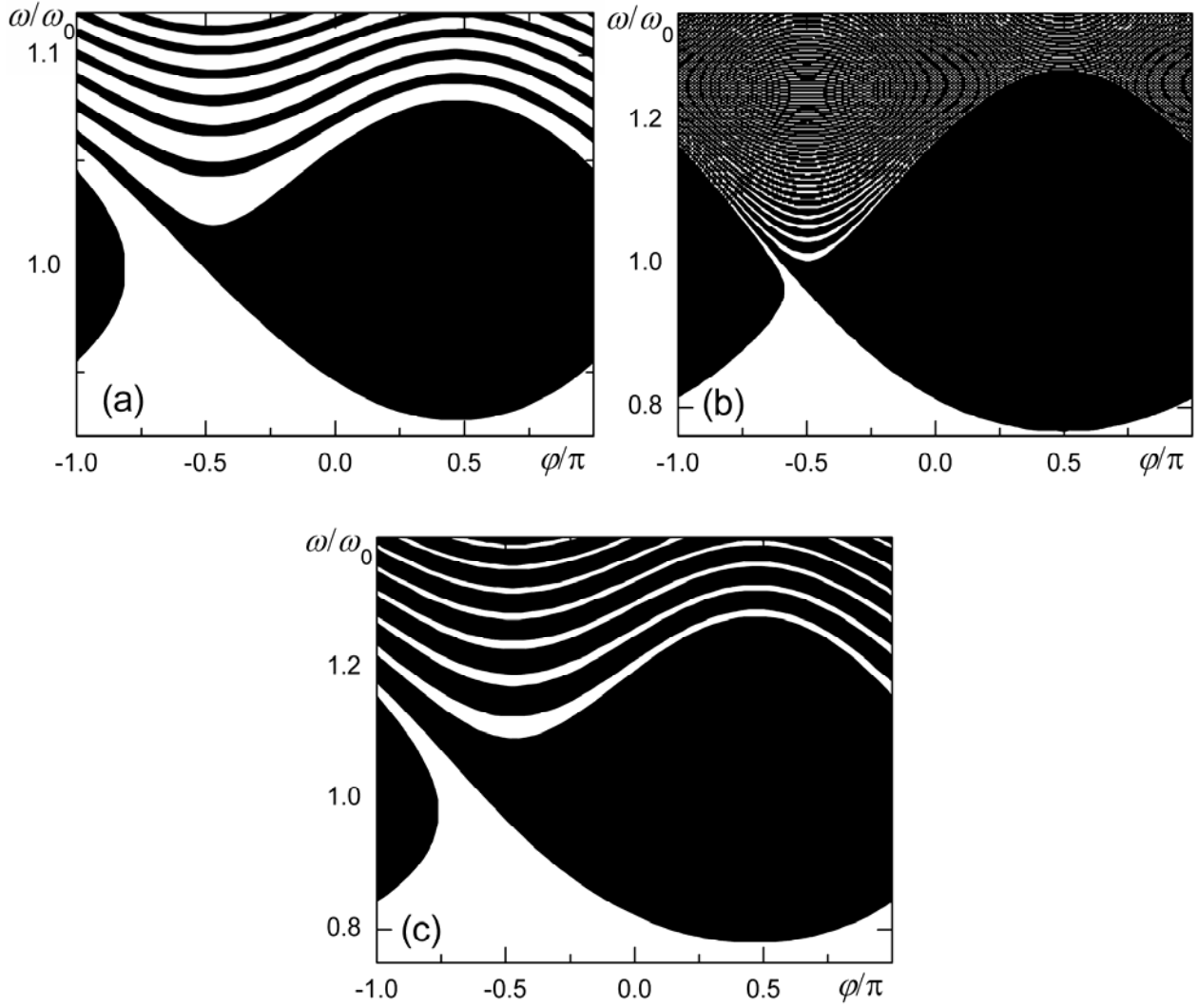

FIG. 7. Calculated parameters of the control force $\omega/\omega_0$ and $\varphi$ (shown in black) for which the stationary operation mode is possible. The calculations were performed for the "long nanotube" case of the phenomenological model. (a) $F_0/F_{0c} = 10$, $Q = 500$; (b) $F_0/F_{0c} = 100$, $Q = 500$; (c) $F_0/F_{0c} = 10$, $Q = 50$.

In both "short nanotube" and "long nanotube" cases of the model we obtained similar results (see FIG. 6). So, it was demonstrated that the choice of the potential for the model has negligible influence on the behavior of the considered oscillator and the simpler "long nanotube" case (disregarding the shape of the dependence of the interwall van der Waals energy on the displacement of the movable wall) can be used for qualitative studies of the oscillator operation for nanotubes longer than 3 nm and oscillation amplitudes above 1 nm. This statement is also confirmed by the results obtained in Section IV with account of the thermodynamic fluctuations.

The relaxation to the stationary operation mode was studied for $\varphi = 0$ and $\omega = \omega_0$. If the amplitude of the control force exceeds the critical value, the stationary operation mode is attained in 10 – 100 ns. During the relaxation to the stationary operation mode, the amplitude and frequency of the NEMS oscillate and tend to their stationary values. The frequency of the gigahertz oscillator approaches the frequency of the control force. The time dependence of the frequency of the gigahertz oscillator can be approximated by

$$v(t) = \frac{\omega}{2\pi} + \frac{\delta\omega}{2\pi}\exp\left(-\frac{t}{\tau_{rel}}\right)\sin\left(\frac{2\pi t}{T_{osc}} + \phi_\omega\right), \qquad (15)$$

where $\tau_{rel}$ is the characteristic time required to reach the stationary operation mode, $\delta\omega$ is the parameter characterizing the magnitude of the frequency oscillations, $T_{osc}$ is the period of the frequency and amplitude oscillations, and $\phi_\omega$ is the fitting parameter. A similar time dependence can be written for the amplitude of the gigahertz oscillator. Note that the variation of the oscillation amplitude revealed by the phenomenological model conforms to the results obtained in the MD simulations (see FIG. 4a,b).

The time parameters characterizing relaxation to the stationary operation mode as functions of the Q-factor and the amplitude of the control force are shown in FIG. 8 and FIG. 9. Note that the period of the frequency and amplitude oscillations exceeds the oscillation period of the gigahertz oscillator by several orders of magnitude. As it is seen in FIG. 8, an increase in the Q-factor leads to an increase in the characteristic time $\tau_{rel} \propto Q^{1.00\pm0.01}$ and the period $T_{osc} \propto Q^{0.50\pm0.01}$, and also in a decrease in the magnitude of the frequency oscillations $\delta\omega \propto Q^{-0.50\pm0.01}$. Near the critical value of the control force amplitude $F_0 = F_{0c}$ (see Eq. (1)), the parameter $\delta\omega$ characterizing the magnitude of the frequency oscillations and the period $T_{osc}$ of the frequency and amplitude oscillations as functions of $\Delta F = F_0 - F_{0c}$ are described by "universal" dependences: $\delta\omega \propto \Delta F^{0.75\pm0.01}$ and $T_{osc} \propto \Delta F^{-0.25\pm0.03}$ (see

FIG. 9). In the stationary operation mode, the frequency and amplitude of the gigahertz oscillator and also the phase shift between the control force and relative velocity of the walls are constant.

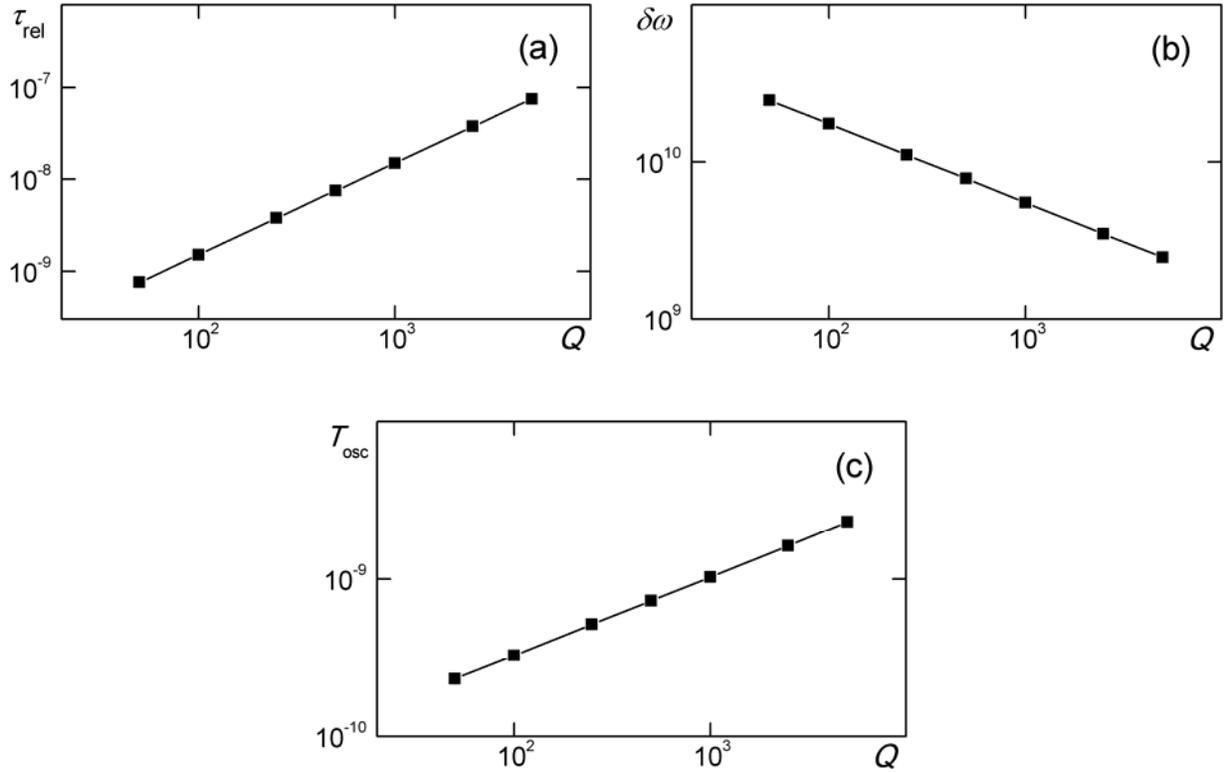

FIG. 8. Calculated (a) characteristic time $\tau_{rel}$ (in s) required to reach the stationary operation mode, (b) magnitude of the frequency oscillations $\delta\omega$ (in s$^{-1}$) and (c) period of the frequency oscillations $T_{osc}$ (in s) as functions of the Q-factor for $F_0 / F_{0c} = 2$, $\omega = \omega_0$ and $\varphi = 0$.

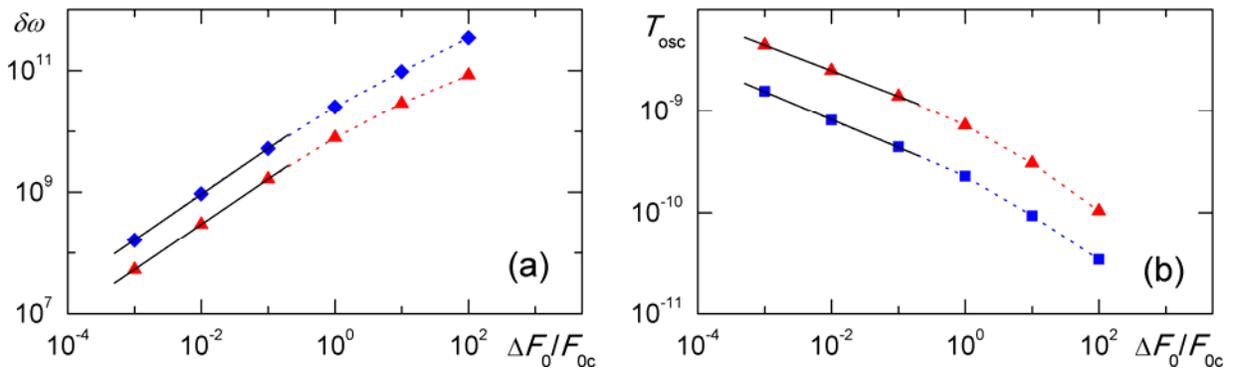

FIG. 9. (Color online) Calculated (a) magnitude of the frequency oscillations $\delta\omega$ (in s$^{-1}$) and (b) period of the frequency oscillations $T_{osc}$ (in s) as functions of $\Delta F_0 / F_{0c} = (F_0 - F_{0c}) / F_{0c}$ for $\omega = \omega_0$, $\varphi = 0$, (■)

$Q = 50$ and (▲) $Q = 500$. Solid straight lines show the approximations of these functions for $\Delta F_0 / F_{0c} \ll 1$.

The obtained dependences of $\delta \omega$ and $T_{osc}$ on the Q-factor and the control force amplitude can be explained in the following way. In the stationary oscillation mode, the phase shift $\varphi_0$ between the control force and the velocity of the movable wall is given by $\cos\varphi_0 = F_{0c} / F_0$. So if $\Delta F / F_{0c} \ll 1$, $\varphi_0 \approx \sqrt{2\Delta F / F_{0c}}$. If the initial phase shift of the control force equals zero ($\varphi = 0$), the number of the oscillation periods needed to reach the phase shift $\varphi_0$ is $N = \varphi_0 \omega / (2\pi \delta \omega)$. During this time, the energy perturbation $\delta E = 2 F_W s \delta \omega / \omega$ is compensated by the excessive work of the control force $N \Delta F s$. So, using Eq. (1) for $F_W$, one gets

$$\frac{\delta \omega}{\omega} = \frac{1}{16}\sqrt{\frac{2\pi \Delta F}{F_{0c}} \frac{\varphi_0}{Q}} \approx \frac{2^{3/4} \pi^{1/2}}{16 Q^{1/2}} \left(\frac{\Delta F}{F_{0c}}\right)^{3/4}. \tag{16}$$

The period of the frequency and amplitude oscillations is given by

$$\frac{T_{osc} \omega}{2\pi} = 4N = \frac{2\varphi_0}{\pi} \frac{\omega}{\delta \omega} \approx 16 \frac{2^{3/4} Q^{1/2}}{\pi^{3/2}} \left(\frac{\Delta F}{F_{0c}}\right)^{-1/4}. \tag{17}$$

Note that in the case $\Delta F / F_{0c} \gg 1$, the phase shift of the control force in the stationary operation mode is $\varphi_0 \approx \pi/2$, and

$$\frac{\delta \omega}{\omega} \approx \frac{\pi}{16 Q^{1/2}} \left(\frac{\Delta F}{F_{0c}}\right)^{1/2}, \qquad \frac{T_{osc} \omega}{2\pi} \approx \frac{16 Q^{1/2}}{\pi} \left(\frac{\Delta F}{F_{0c}}\right)^{-1/2}. \tag{18}$$

If the initial phase shift of the control force considerably differs from zero ($\varphi \sim \pi$), the stationary operation mode is possible only in the case $\Delta F / F_{0c} \gg 1$, and the dependences (18) are still valid. Note that estimates (16) – (18) not only give the same qualitative dependences as presented in FIG. 8 and FIG. 9, but also the numerical factors with an accuracy of 10-50 %.

**IV. THERMODYNAMIC FLUCTUATIONS IN GIGAHERTZ OSCILLATOR**

As it was shown in Sec. II using the MD simulations, thermodynamic fluctuations lead to the breakdown of the stationary oscillation. To examine the effect of fluctuations with the help of the one-dimensional model, one should introduce noise $\xi(t)$ into motion equation (11). For the (5,5)@(10,10) nanotube-based oscillator, the relative deviation $\delta_E$ of the relative energy change $\Delta E/E$ over the half-period of the oscillation calculated through the MD simulations was shown to be in good agreement with the analytical results obtained on the basis of the fluctuation-dissipation theorem. So, thermal noise at temperature $T$ should be a Gaussian white noise of zero mean $\langle \xi(t) \rangle = 0$ which satisfies the fluctuation-dissipation relation[53,54] $\langle \xi(t)\xi(t-\tau) \rangle = 2\eta k_B T \delta(\tau)$. Here $\delta(\tau)$ is the Dirac delta-function. Integrating motion equation (11), we changed the value $\xi(t)$ randomly with the time step $\tau_\xi$. Therefore, the dispersion of $\xi(t)$ was determined by

$$\langle \xi^2 \rangle = \frac{2\eta k_B T}{\tau_\xi}. \tag{19}$$

Eq. (19) can be presented in the obvious form

$$\langle \xi^2 \rangle = \delta^2 \frac{T_s}{2\tau_\xi} \langle |F_{\mathrm{fr}}| \rangle_s^2, \tag{20}$$

where $T_s$ is the period of the oscillation with the amplitude $s$, $\langle |F_{\mathrm{fr}}| \rangle_s = 4\eta s / T_s$ is the average friction force for the oscillation with the amplitude $s$, and $\delta = \sqrt{T_s k_B T / (4\eta s^2)}$ is the coefficient characterizing the ratio of the noise to the friction force. In the "long nanotube" case, as it follows from expression (13) for the friction coefficient $\eta$ and expression (14) for the oscillation period, $\delta$ is given by the equation

$$\delta = \sqrt{\frac{16}{3} \frac{k_B T Q}{E}}. \tag{21}$$

Comparing Eq. (7) and Eq. (21), one can see that $\delta / \delta_E \approx 1$. The point is that $\delta$ is determined by the ratio of the noise to the friction force averaged over time, while $\delta_E$ is given by the ratio of the same quantities averaged over the displacements of the movable wall. So, $\delta$ and $\delta_E$ should differ only by a numerical factor of an order of 1.

If the fluctuation-dissipation relation is not satisfied, one can suppose that the noise $\xi(t)$ is white and introduce it according to Eq. (20). However, in this case, the dispersion of $\xi(t)$ is not related to the Q-factor and temperature and can be extracted, for example, from the values of $\delta_E$ obtained by MD simulations.

In our simulations, we changed the value $\xi(t)$ randomly with the time step $\tau_\xi$ equal to the simulation time step 1 fs in the numerical calculations for the "long nanotube" and "short nanotube" cases and equal to the oscillation half-period in the semi-analytical calculations for the "long nanotube" case. The dispersion of $\xi(t)$ was assumed to be constant. We characterize this dispersion by the value of $\delta$ in Eq.(20) corresponding to the initial oscillation amplitude 1 nm. Note that the values of $\delta$ were taken of the order of $\delta_E$ calculated through the MD simulations (see TABLE I). The breakdown of the stationary mode was defined as the moment when the oscillation amplitude becomes less than 0.4 nm. The lifetime of the stationary mode $\tau_s$ (i.e., the time to the breakdown of the controlled oscillation) for given $F_0 / F_{0c}$ and $\delta$ was averaged over 100 – 400 numerical solutions of Eq. (11).

In both "short nanotube" and "long nanotube" cases, the model predicts that the fluctuations actually lead to the breakdown of oscillations, i.e., the average lifetime of the stationary mode is finite. The calculations with the one-dimensional model have shown that the lifetime $\tau_s$ increases with decreasing $\delta$ (see FIG. 10) and approaches infinity at $\delta \to 0$. The lifetime of the stationary mode can also be increased with increasing the amplitude of the control force (see FIG. 10). More specifically, the lifetime almost exponentially depends on the ratio $F_0 / F_{0c}$ and, for $\delta = 1.5$, becomes longer than 1 ms at $F_0 / F_{0c} > 10$. The explanation of the dependences obtained is given below. Note that the model predicts the average lifetime of the stationary mode to be about 1 ns at the amplitude of the control force slightly above the critical value, in good agreement with the breakdown of the stationary mode observed in the MD simulations (see FIG. 4c). The curves obtained in the "short nanotube" and "long nanotube" cases of the model are very close to each other, thus proving once again the applicability of the "long nanotube" case of the model (see FIG. 10). The semi-analytical approach with the time step $\tau_\xi$ equal to the oscillation half-period provides shorter lifetimes compared to the calculations with the time step $\tau_\xi = 1\,\text{fs}$ (see FIG. 10). However, the qualitative dependences of the lifetime on $F_0/F_{0c}$ and $\delta$ are identical.

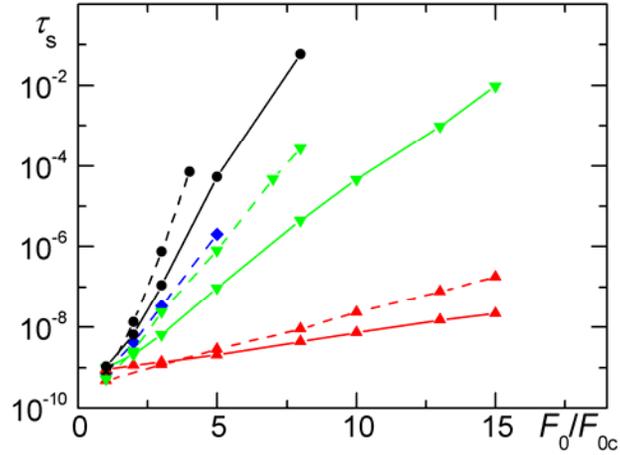

FIG. 10. (Color online) Calculated lifetime $\tau_s$ (in s) of the stationary operation mode as a function of $F_0/F_{0c}$ for $Q = 55$, $\varphi = 0$ and $\omega = \omega_0$. Dashed lines show the results of the numerical calculations with $\tau_\xi = 1\,\text{fs}$: in the "long nanotube" case (▲) $\delta = 3$, (▼) $\delta = 1.5$, (●) $\delta = 1$; in the "short nanotube" case (♦) $\delta = 1.5$. Solid lines show the results of the semi-analytical calculations for the "long nanotube" case with $\tau_\xi$ equal to the half-period of the oscillation.

To explain the calculated dependence of the lifetime of the stationary operation mode of the gigahertz oscillator on $F_0/F_{0c}$ and $\delta$, we performed the analysis of the oscillation energy distribution function in the "long nanotube" case. Our analysis was based on the following assumptions: (1) the fluctuation-dissipation relation is satisfied and (2) friction coefficient $\eta$ is constant and independent of the oscillation amplitude. However, we believe that the qualitative conclusions made in the analysis below are valid for wider conditions of NEMS operation. The oscillation energy distribution function $f(E,t)$ characterizes the probability for the oscillator to have oscillation energy $E$ at time instance $t$. The evolution of the oscillation energy distribution function in the gigahertz oscillator is determined by the Fokker–Planck equation

$$\frac{\partial f}{\partial t} = \frac{\partial}{\partial E}\left(D\frac{\partial f}{\partial E} - uf\right). \tag{22}$$

Let us first consider the free oscillation. The damping (or drift) of the oscillation energy is determined as

$$u = \frac{dE}{dt} = -\frac{E}{QT_s}. \tag{23}$$

Note that the product of the Q-factor and the oscillation period $QT_s$ is independent of the oscillation amplitude for the considered phenomenological model with the constant friction coefficient $\eta$ (see Eq.(13)).

Based on the fluctuation-dissipation theorem, it was shown that the thermal noise leads to the diffusion of the oscillation energy

$$\langle(\delta E)^2\rangle = 2Dt \tag{24}$$

with the diffusion coefficient

$$D = \alpha\frac{k_{\mathrm{B}}TE}{QT_s}, \tag{25}$$

where $\alpha \approx 1$ is a numerical factor found below using the phenomenological model.

Thus, the Fokker-Planck equation takes the form

$$\frac{\partial f}{\partial t} = \frac{\partial}{\partial E}\left(aE\frac{\partial f}{\partial E} + bEf\right), \tag{26}$$

where

$$a = \alpha \frac{k_{\mathrm{B}} T}{Q T_s}, \qquad b = \frac{1}{Q T_s}. \tag{27}$$

To solve Eq. (26), let us introduce the variables

$$\zeta = E \exp(bt) - E_0, \qquad \chi = \exp(bt) - 1, \tag{28}$$

where $E_0$ is the initial oscillation energy.

For these variables and $F(\zeta, \chi) = f(E, t) \exp(-bt)$, Eq. (26) is reduced to

$$\frac{\partial F}{\partial \chi} = \frac{a}{b} \frac{\partial}{\partial \zeta} \left( (\zeta + E_0) \frac{\partial F}{\partial \zeta} \right). \tag{29}$$

In the case $|\zeta| \ll E_0$, which means that the width of the oscillation energy distribution is small compared to the current average oscillation energy, one gets a simple one-dimensional diffusion equation

$$\frac{\partial F}{\partial \chi} = \frac{a E_0}{b} \frac{\partial^2 F}{\partial \zeta^2}. \tag{30}$$

From these considerations it can be seen that the solution of Eq. (26) for the initial condition $f(E, 0) = \delta(E - E_0)$ can be approximated as

$$f(E, t) \approx \sqrt{\frac{b \exp(2bt)}{4 \pi a E_0 (\exp(bt) - 1)}} \exp\left( -\frac{b}{4 a E_0} \frac{(E - E_0 \exp(-bt))^2}{(\exp(bt) - 1)} \exp(2bt) \right) \tag{31}$$

for $t \ll T_s Q \ln(E_0 / k_{\mathrm{B}} T)$. In the case of the small damping ($t \ll T_s Q$), Eq. (31) takes the simple form

$$f(E, t) \approx \frac{1}{\sqrt{4 \pi a E_0 t}} \exp\left( -\frac{(E - E_0 (1 - bt))^2}{4 a E_0 t} \right). \tag{32}$$

For the considered phenomenological model with the constant dispersion of the noise $\langle \xi^2 \rangle$, we used Eq. (19) to express the diffusion coefficient $D$ and the parameter $a$ in terms of the parameters of the model

$$D = \alpha \frac{\langle \xi^2 \rangle \tau_\xi E}{3m} = \frac{3\alpha}{16} \frac{\delta^2 E}{Q} \frac{E}{QT_s}, \quad a = \frac{3\alpha}{16} \frac{\delta^2 E}{Q^2 T_s}. \tag{33}$$

Note that the diffusion coefficient still linearly depends on the oscillation energy.

We numerically calculated the energy distribution function for damping oscillations in the "long nanotube" case with $\tau_\xi = 1$ fs as a function of time (see FIG. 11). To fit the energy distribution function, we estimated numerical factor $\alpha$ to be $\alpha = 0.75$ in Eqs. (25), (27), (33). As is seen in FIG. 11, solution (31) for this value of $\alpha$ is in good agreement with the results of the numerical calculations.

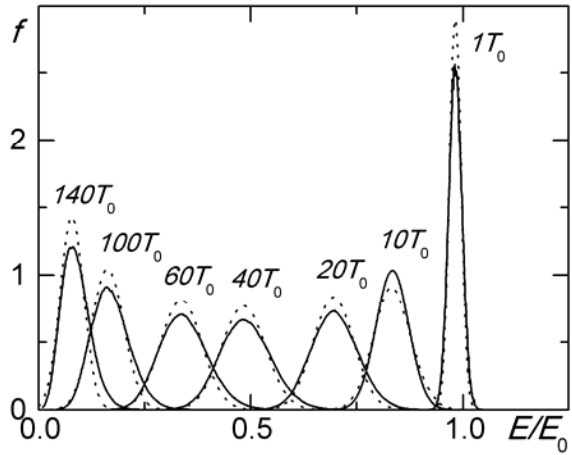

FIG. 11. (Color online) Energy distribution function $f(E/E_0)$ for the damping oscillation at different time instances for $Q = 55$ and $\delta = 1.5$. Solid lines show the results of the numerical calculations in the "long nanotube" case with $\tau_\xi = 1$ fs. Corresponding functions calculated using Eq. (31) are shown with the dotted lines.

In the stationary operation mode, the oscillation energy drift can be determined as the rate of energy relaxation to the stationary value $E_s$. As it was shown above by the MD simulations and calculations with the phenomenological model, if the oscillation energy $E$ is given a small perturbation, energy oscillations occur with the period $T_{osc} \propto T_s \sqrt{Q}$, which is considerably greater than the oscillation period $T_s$ (see FIG. 8c). Nevertheless, since the period of the energy oscillations $T_{osc}$ is much shorter than the characteristic time $\tau_{\text{rel}}$ required for the reversion to the stationary operation mode, we believe that one can neglect the energy oscillations and approximate the drift term in the Fokker–Planck equation (22) by the relaxation rate of the absolute value of the oscillation energy deviation from the stationary value $|E - E_s|$ averaged over the time $T_{\text{osc}}$. The time of relaxation $\tau_{\text{rel}}$ to the stationary operation mode was shown to linearly depend on the Q-factor $\tau_{\text{rel}} \approx T_s Q$ (see FIG. 8a). Thus, the drift term can be approximated as

$$u = \frac{d \langle |E - E_s| \rangle_{T_{\text{osc}}}}{dt} \approx -\frac{\langle |E - E_s| \rangle_{T_{\text{osc}}}}{Q T_s}. \tag{34}$$

So, the Fokker–Planck equation for the energy distribution function in the stationary operation mode takes the form

$$\frac{\partial f}{\partial t} = \frac{\partial}{\partial E}\left( aE \frac{\partial f}{\partial E} + b(E - E_s) f \right), \tag{35}$$

Therefore, the distribution function in the stationary operation mode, which is established within the time of about $\tau_{\text{rel}}$ is given by

$$f(E) = C \exp\left( -\int_{E_s}^{E} \frac{b(E' - E_s)}{aE'} dE' \right) = C \exp\left( -\frac{b}{a}\left( E - E_s - E_s \ln\left(\frac{E}{E_s}\right) \right) \right), \tag{36}$$

where $C$ is a constant determined by the normalization of $f(E)$. For $|E - E_s| \ll E_s$, one gets

$$f(E) = C \exp\left( -\frac{bE_s}{2a}\left( \frac{E - E_s}{E_s} \right)^2 \right). \tag{37}$$

Substituting expressions (27) for $a$ (with $\alpha = 0.75$) and $b$, one gets

$$f(E) = C \exp\left( -\frac{E_s}{2\alpha k_{\text{B}} T}\left( \frac{E - E_s}{E_s} \right)^2 \right). \tag{38}$$

It can be seen that the width of the oscillation energy distribution depends on the ratio of the stationary oscillation energy $E_s$ to the thermal kinetic energy $k_{\text{B}} T$ and is independent of the Q-factor.

In terms of the parameters of the considered phenomenological model with the constant noise dispersion, the oscillation energy distribution function in the stationary operation mode is obtained by substituting Eq. (27) for $b$ and Eq. (33) for $a$ into Eq. (37)

$$f(E) = C \exp\left(-\frac{8Q}{3\alpha\delta^2}\left(\frac{E-E_s}{E_s}\right)^2\right). \tag{39}$$

Using the phenomenological model, we numerically calculated the oscillation energy distribution function in the stationary operation mode (see FIG. 12) averaged over 50 simulations for each set of parameters: Q-factor $Q$, amplitude of the control force $F_0$, and the ratio $\delta$ of the noise to the friction force. It can be seen in FIG. 12 that there is a critical value of oscillation energy $E_c$ which separates two operation modes of the gigahertz oscillator: the damping oscillation for $E < E_c$ and the stationary operation mode where operation can be controlled at $E > E_c$. The critical value $E_c$ can be found as the point where the derivative of the oscillation energy distribution function has a jump. The results of the numerical calculations are in reasonable agreement with Eq. (39) for $E > E_c$ (see FIG. 12), so neglecting the energy oscillations in the estimation of the drift term in the Fokker–Planck equation is adequate for the qualitative description of the system behavior.

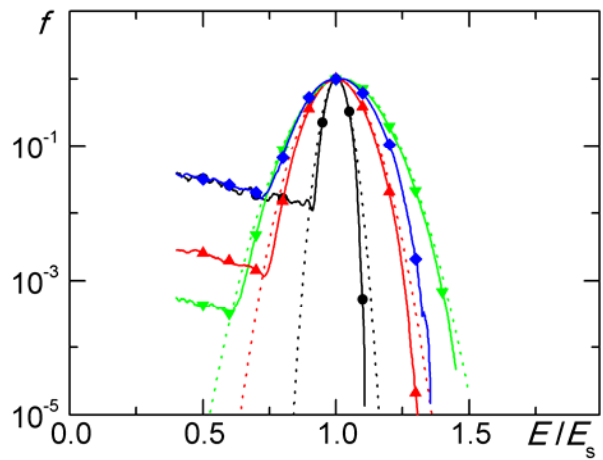

FIG. 12. (Color online) Energy distribution functions $f(E/E_s)$ for the controlled oscillations at $\omega = \omega_0$ and $\varphi = 0$. Solid lines show the results of the numerical calculations in the "long nanotube" case with

$\tau_\xi = 1$ fs: (♦) $\delta = 2.0$, $Q = 55$, $F_0 / F_{0c} = 4$; (▼) $\delta = 2.0$, $Q = 55$, $F_0 / F_{0c} = 8$; (▲) $\delta = 1.5$, $Q = 55$, $F_0 / F_{0c} = 4$; (●) $\delta = 2.0$, $Q = 500$, $F_0 / F_{0c} = 4$. Corresponding functions calculated using Eq. (39) are shown with the dotted lines.

The probability that $E$ takes some critical value $E_c$, which results in the breakdown of the stationary oscillation, is proportional to $f(E_c)$. Therefore, the lifetime of the stationary oscillation is determined by

$$\tau_s \approx T_s Q \exp\left( \frac{8Q}{3\alpha \delta^2} \left( \frac{E_c - E_s}{E_s} \right)^2 \right), \quad (40)$$

where the pre-exponential factor is chosen so as to provide $\tau_s \approx T_s Q$ for $F_0 / F_{0c} = 1$.

The critical value $E_c$ can be estimated from the following considerations. For the work of the control force to compensate the deviation of the oscillation energy $E - E_s$ and thus to stabilize the oscillation, at least $N \approx (E - E_s)/(2(F_0 - F_{0c})s)$ periods of the oscillation are required. However, within this time, a phase shift $N\omega_0 \Delta T_s$ between the control force and the oscillation is accumulated, where $\Delta T_s = T_s (E - E_s)/(2E_s)$ is the change of the oscillation period due to the fluctuation of the oscillation amplitude, and the actual work of the control force is less than $2N(F_0 - F_{0c})s$. If the accumulated phase shift is significant $N\omega_0 \Delta T_s \geq \varphi_c$, the actual work of the control force within the considered $N$ periods is too small to compensate the dissipation. In this case, the damping of the oscillation should occur. Substituting expressions for $N$ and $\Delta T_s$ into the equation $N\omega_0 \Delta T_s \approx \varphi_c$, one gets that the critical value $E_c$ of the oscillation energy is approximately determined by

$$\left( \frac{E_c - E_s}{E_s} \right)^2 \approx \frac{2\varphi_c}{\pi} \frac{F_0 - F_{0c}}{F_W} = \frac{\pi \varphi_c}{16} \frac{F_0 - F_{0c}}{QF_{0c}}. \quad (41)$$

Substituting Eq. (41) into Eq. (40), the lifetime of the stationary oscillation is roughly given by

$$\tau_s \approx T_s Q \exp\left(\frac{\pi \varphi_c}{6\alpha} \frac{1}{\delta^2}\left(\frac{F_0}{F_{0c}} - 1\right)\right). \tag{42}$$

For $F_0 / F_{0c} \gg 1$, the phase shift $\varphi_0$ between the control force and the velocity of the movable wall in the stationary operation mode tends to $\pi/2$; therefore, the critical phase shift $\varphi_c$ tends to $\pi$. The obtained expression (42) for the lifetime of the stationary operation mode is in qualitative agreement with the calculated dependence of the lifetime on the control force amplitude (see FIG. 10).

On the basis of Eq. (42) and FIG. 10, let us analyze the possibility to increase the lifetime of the stationary operation mode $\tau_s$ of the oscillator at a given oscillation amplitude and frequency using the external control parameters. In accordance with Eq. (42) and FIG. 10, the lifetime of the stationary operation mode $\tau_s$ exponentially depends on the amplitude of the control force $F_0$ and, therefore, can be increased significantly by increasing the amplitude of the control force $F_0$. In the case where the fluctuation-dissipation theorem is valid and at the given oscillation amplitude and frequency, $\delta$ depends only on temperature (see Eq. (21)). However, due to the Q-factor being almost inversely proportional to temperature (see TABLE I), $\delta$ can only slightly be changed with temperature. As a result, the lifetime of the stationary operation mode $\tau_s$ also weakly depends on temperature.

The critical value $\nu_c$ of frequency corresponding to the critical value $E_c$ of the oscillation energy is determined by the relation

$$\frac{\nu_c - \nu}{\nu} = \frac{\Delta \nu_c}{\nu} = \frac{E_c - E_s}{2 E_s} \tag{43}$$

where $\nu = \omega/(2\pi)$ is the frequency of the control force. From Eqs. (41) and (42) it follows that $\Delta \nu_c / \nu \propto (F_0 / F_{0c})^{1/2}$ for $F_0 / F_{0c} \gg 1$. Such a dependence of the critical frequency on the control force amplitude correlates with the square-root like dependence for the lower limit of the ratio $\omega/\omega_0$ on $F_0 / F_{0c}$ presented in FIG. 6a,b. This is because, in the both cases, we consider the maximum relative difference between the oscillation frequency and the frequency of the control force for which the oscillation is still sustained. Note that estimates of the oscillator stability regions similar to Eqs. (41) and (43) should be valid for any anharmonic oscillator with a strong dependence of the oscillation period on the oscillation energy $dT_s/dE \sim T_s/E$. Therefore, the lifetime of the stationary operation mode should be finite for any strongly anharmonic oscillator.

Up to now atomistic simulation of nanotube-based oscillators was restricted to sizes of hundreds of nanometers and simulation times of tens of nanoseconds[23,24,25,26,27,28,29,30,31,32,33,34,35,36,37,39]. The proposed phenomenological model taking into account thermodynamic fluctuations makes it possible to extend the simulation time up to 1 ms. Let us discuss the possibility to use this model not only for long simulation times but also for sizes of the oscillator greater than 3 nm.

The proposed model based on the motion equation (11) has no size restrictions and can be applied to any size of the oscillator in the case if: 1) there are no resonances between the telescopic oscillation and other vibrational modes of the system; 2) the phenomenological parameters for the random noise force and friction force are available.

Let us analyze the restrictions imposed on the nanotube length by condition (1). Among the low-frequency nanotube vibrational modes, there are long-wave acoustic vibrational modes[57,58,59,60], squash modes[61] and relative vibrational modes of the walls[62,63]. According to our previous calculations, for the (5,5)@(10,10) nanotube of 3 nm length, the fundamental frequency of the doubly degenerate transverse acoustic modes is 1.5 THz, the fundamental frequency of the squash mode is about 1 THz and the frequencies of the non-axial translational relative vibrations of the inner wall inside the outer wall are 0.25-0.33 THz. For this nanotube, there is no resonance between the telescopic oscillation and any other nanotube vibrational modes. With increasing the nanotube length, the fundamental frequencies of the longitudinal and torsional acoustic modes decrease as $\propto 1/L$ [57,58,60] (where $L$ is the nanotube length), while the frequencies of the relative vibrations of the nanotube walls and of the squash modes weakly depend on the nanotube length. Since the frequency of the telescopic oscillation is inversely proportional to the nanotube length (see Eq. (14) for the case $s \propto L$), these modes do not become resonant with the telescopic oscillation. However, from the Euler-Bernoulli beam theory[64] it follows that the fundamental frequency of the transverse acoustic (flexural) vibrational modes should decrease with increasing the nanotube length faster than the frequency of the telescopic oscillation

$$f_{tr}^0 = \frac{22.4}{2\pi L^2}\sqrt{\frac{YI}{\rho A}}, \qquad (44)$$

where $Y \approx 1\,\text{TPa}$ is the nanotube Young's modulus[65,66,67], $\rho \approx 2.2 \cdot 10^3 \text{ kg/m}^3$ is the density of carbon atoms, $I \approx \pi R^4$ is the nanotube areal moment of inertia, $A \approx \pi R^2$ is the nanotube cross-sectional area and $R$ is the outer wall radius. From the equality of the fundamental frequency of the transverse acoustic modes (44) and the frequency of the telescopic oscillation (see Eq. (14) for the case $s = L/3$), the nanotube length at which the resonance between these modes is possible can be estimated as

$$\frac{L_{max}}{R} \approx 10.3R\sqrt{\frac{Y}{F_W}} \approx 200. \qquad (45)$$

So for the (5,5)@(10,10) nanotube, the maximum length up to which the one-dimensional model is applicable is about $L_{max} \approx 140\,\text{nm}$. For longer nanotubes, coupling of the telescopic oscillation with the transverse acoustic modes should be taken into account.

In the present paper we only considered the case when there is no resonance between the telescopic oscillation and other nanotube vibrational modes. As it was shown by MD simulations, in the case of the resonance, not only the Q-factor of the oscillator decreases drastically but the level of thermodynamic fluctuations is also a few times greater than that under the conditions when the fluctuation-dissipation relation is satisfied (see Eq. (7)). Therefore, according to Eq. (42), the lifetime of the stationary operation mode in the case of the resonance should be significantly smaller compared to that for the case when the resonance is absent. So the operation of the oscillator in the case when there is a resonance with any nanotube vibrational mode should be avoided.

As for condition (2), the phenomenological parameters for the short nanotube of 3 nm length considered above were extracted from the MD simulations and validity of the model was based on matching to the MD results. For longer nanotubes, the phenomenological parameters of the model can be determined accurately on the basis of experimental measurements. Note that as the fluctuation-dissipation theorem, which relates the thermal noise to the energy dissipation rate, is valid for these nanotubes, it is sufficient to measure only the Q-factor of the oscillator.

On the basis of the obtained results, let us discuss the restrictions imposed by thermodynamic fluctuations on sizes of the nanotube-based NEMS for which control over the NEMS operation is possible. From Eq. (42) and FIG. 10, it is seen that the lifetime of the stationary operation mode $\tau_s$ can be increased by decreasing the level of fluctuations $\delta$. According to Eq. (21), $\delta$ is proportional to the square root of the Q-factor divided by the oscillation energy. Therefore, a decrease of $\delta$ can be achieved by an increase of the oscillation amplitude. However, for the oscillation amplitude greater than about 30% of the inner wall length, the dissipation rate strongly increases (and, consequently, the Q-factor strongly decreases) with an increase of the oscillation amplitude due to the excitation of low-frequency vibrational modes[26,35]. Thus, for an oscillator with a certain length the minimum level of fluctuations $\delta$ and, therefore, the maximum lifetime of the stationary operation mode $\tau_s$ correspond to some optimal oscillation amplitude about 30% of the oscillator length. From the MD simulations it follows that if the ratio of the initial telescopic extension of the inner wall to the oscillator length is maintained, the Q-factor is independent of the oscillator length (in the region $L << L_{max}$, see Eq. (45)). In the same case, the oscillation energy is proportional to the oscillator length. Hence, one gets the minimum value of $\delta$ for oscillators with the optimal oscillation amplitude to be inversely proportional to the square root of the oscillator length. In accordance with Eq. (42), this means that the lifetime of the stationary operation mode $\tau_s$ for such oscillators increases exponentially with increasing their length. For example, based on the data of TABLE I for the 3.1 nm (5,5)@(10,10) nanotube-based oscillator, one can estimate that, for the oscillator longer than 35 nm, $\delta$ should be below 0.4. For $F_0 / F_{0c} = 2$, the lifetime of the stationary operation mode would already exceed 1 s. Thus, by the example of the gigahertz oscillator, we demonstrated that thermodynamic fluctuations can impose crucial restrictions on sizes of NEMS for which control of the NEMS operation is possible.

## V. CONCLUSION

In summary, we performed molecular dynamics (MD) simulations of the controlled operation of the nanotube-based gigahertz oscillator. The feasibility of control over a movable nanotube wall which is functionalized so that it has an electric dipole moment by a nonuniform electric field was demonstrated. The MD simulations were used to obtain the oscillator Q-factor and the characteristics of the thermal noise. To study the possibility of the stationary operation mode (operation mode with a constant frequency) at a simulation time of 1 ms, a phenomenological one-dimensional model with the parameters derived from MD simulations was proposed. Using this model, the control force parameters (amplitude $F_0$, angular frequency $\omega$, and initial phase shift $\varphi$ between the control force and the velocity of the movable wall) at which the stationary operation mode is possible were determined. In particular, it was shown that the ranges of $\omega$ and $\varphi$ which correspond to the stationary operation mode become wider with increasing the amplitude of the control force.

Significant thermodynamic fluctuations in the nanotube-based gigahertz oscillator were observed by the MD simulations of damping oscillations. The multi-scale simulations (both MD simulations and simulations with the phenomenological model) revealed that the fluctuations cause the breakdown of the stationary operation mode. The average lifetime of the stationary operation mode was found to decrease with increasing the level of fluctuations or decreasing the amplitude of the control force. These dependences were explained through the analysis of the oscillation energy distribution function calculated on the basis of the Fokker–Planck equation. The investigations performed showed that the thermodynamic fluctuations impose restrictions on the sizes of NEMS for which the control of the NEMS operation is feasible.

## ACKNOWLEDGEMENTS

This study was supported by the Belarussian–Russian Foundation for Basic Research within the framework of the joint program BRFFI–RFFI (Grant F10R-062) and the Russian Foundation for Basic Research (Grants 08-02-00685 and 10-02-90021-Bel).